\documentclass[aps,prb,twocolumn,notitlepage,showpacs,floatfix,superscriptaddress,groupedaddress,superscriptaddress]{revtex4-1}

\usepackage{times,amsmath,amsfonts,amssymb,mathrsfs,graphics,graphicx,color,comment,bm}
\usepackage[next]{inputenc}
\usepackage[dvips]{epsfig}
\usepackage[pdfstartview=FitH]{hyperref}
\usepackage{natbib}
\usepackage{pdfsync}
\usepackage{appendix}

\begin{document}

\title{Quantum critical phase diagram of bond alternating Ising model with Dzyaloshinskii-Moriya interaction:
signature of ground state fidelity}

\author{N. Amiri}
\address{Department of Physics, Sharif University of Technology, Tehran 11155-9161, Iran}
\author{A. Langari}
\address{Department of Physics, Sharif University of Technology, Tehran 11155-9161, Iran}
\email{langari@sharif.edu}
\homepage{http://spin.cscm.ir}

\begin{abstract}
We present the zero temperature phase diagram of the bond alternating Ising chain in the
presence of Dzyaloshinskii-Moriya interaction. An abrupt change in ground state fidelity
is a signature of quantum phase transition. We obtain the renormalization of fidelity in terms
of quantum renormalization group without the need to know the ground state. We calculate the fidelity
susceptibility and its scaling behavior close to quantum critical point (QCP) to find the 
critical exponent which governs the divergence of correlation length.
The model consists of a long range antiferromagnetic order with nonzero staggered
magnetization  which is separated from a helical ordered phase at QCP.
Our results state that the critical exponent is independent of the bond alternation
parameter ($\lambda$) while the maximum attainable helical order depends on $\lambda$.
\end{abstract}

\maketitle   

\section{Introduction\label{sec1}}
One of the key features of strongly correlated electron systems is their zero temperature behavior
where quantum fluctuations have the dominant role to classify different novel phases which
are separated by quantum phase transitions \cite{vojta}. In this case, ground state (GS) is the sole
candidate which receives drastic changes at QCP driven by some external
parameters. Identification of QCPs is always a challenging task specially when 
quantum correlations diminish a single-particle picture.
Within last couple of years different measures of quantum
entanglement have been proposed to be a new toolkit to detect and categorize QCPs \cite{amico2009}.
Recently, ground state fidelity--the overlap of ground state at two slightly different values
of coupling constants--has attracted intensive attention as a proper quantity to signal
QCP without the need to know the structure of phases close to 
phase transition \cite{zanardi2006,Gu2010}.

An abrupt drop of fidelity in the vicinity of QCP is a consequence of an essential change in the 
structure of GS which is usually accompanied with a divergence of fidelity susceptibility. 
Meanwhile, it is rather difficult to obtain the GS of a strongly correlated many
body system. However, recent successful attempts to calculate some measures of 
entanglement \cite{kargarian}
by quantum renormalization group (QRG) approach \cite{qrg}
lead to a unified formalism \cite{Langari2012}
to obtain fidelity in the thermodynamic limit ($N\rightarrow \infty$) without
knowing the GS of a large system. 
It suggests to implement QRG for obtaining the renormalization
of fidelity for the bond alternating Ising model with Dzyaloshinskii-Moriya (DM) interaction.
DM interaction \cite{Dzyaloshinskii,Moriya} which roots to the spin-orbit coupling is the
antisymmetric super-exchange which leads to helical magnetic structures as the most likely 
candidates to host ferroelectricity \cite{kimura2005}. 

In this article we study the quantum critical behavior of the one dimensional bond alternating
Ising model with DM interaction in terms of renormalization of fidelity. 
The model represents an antiferromagnetic (N\'{e}el) long range order for small DM interaction while
it undergoes via a continuous phase transition to a helical ordered phase. The ground state
in the  N\'{e}el phase is a product state which has essentially zero quantum entanglement 
and nonzero staggered magnetization while
the helical phase poses a correlated quantum ground state with zero staggered magnetization.
Within a classical picture the helical phase could be assumed as slightly rotating spins 
along the direction of chain. In terms of Landau theory of critical phenomena the 
staggered magnetization could be chosen as the proper order parameter.
We have shown that
the divergence in fidelity susceptibility (FS) at the QCP is
an appropriate signature to find QCP which is more pronounced 
than the second derivative of ground state energy \cite{chen2008}.
The presented scheme makes us to find the critical points and its corresponding exponents more accurately.

\section{Quantum renormalization group}

\begin{figure}[t]%
\includegraphics*[width=\linewidth,height=2.7cm]{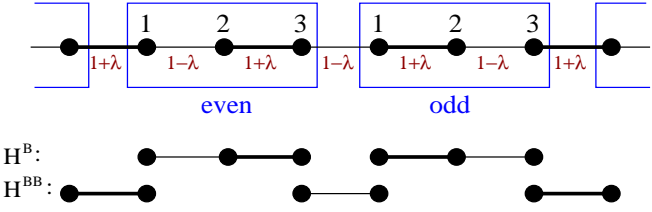}
\caption{Decomposition of the lattice to 3-sites blocks. The intra-block Hamiltonian ($H^B$)
and inter-block one ($H^{BB}$) are represented schematically in the bottom part.}
\label{blocks}
\end{figure}

The Hamiltonian of bond alternating Ising chain with DM
interaction on a periodic chain of N sites is defined
\begin{equation}
\label{eq1}
H=J\sum_{i=1}^{N}\Big[\big(1-(-1)^{i}\lambda \big) S_{i}^{z}S_{i+1}^{z}+\vec D \cdot(\vec S_{i}\times \vec S_{i+1})\Big],
\end{equation}
where $\vec S_{i}$ is the spin-1/2 operator at site $i$, $ \lambda$ describes the relative strength of 
the alternating coupling and $J>0$ shows the nearest-neighboring antiferromagnetic coupling. 
$\vec D$ is the vector of DM interaction which is considered in 
$z$-direction, i.e. $\vec D=D \hat{z}$.
To apply QRG \cite{qrg}, the spin chain is decomposed to three-sites blocks (see Fig.\ref{blocks}) 
where the intra-block Hamiltonian is
$H^B$ and the inter-block one is $H^{BB}$ and their sum defines the whole Hamiltonian, $H=H^B+H^{BB}$
(see the bottom part of Fig.\ref{blocks}) .
In this respect we have  
$H^{B}=\sum_{I=1}^{N/3} h_I^B$, where
\begin{eqnarray}
h_{I}^B=J\sum_{l=1}^2 \Big([1+(-1)^{I+l}\lambda]S_{l,I}^zS_{l+1,I}^z \nonumber \\
+D(S_{l,I}^xS_{l+1,I}^y-S_{l,I}^yS_{l+1,I}^x)\Big),
\end{eqnarray}
and similarly 
$H^{BB}=\sum_{I=1}^{N/3} h_{I,I+1}^{BB}$, 
\begin{eqnarray}
h_{I,I+1}^{BB}=J\Big([1-(-1)^I\lambda]S_{3,I}^zS_{1,I+1}^z \nonumber \\
+D(S_{3,I}^xS_{1,I+1}^y-S_{3,I}^yS_{1,I+1}^x)\Big).
\end{eqnarray}
The block Hamiltonian ($h_I^B$) is diagonalized exactly and the two lowest energy eigenvectors 
($|\Psi_I^{\pm}\rangle$)
are kept to construct the embedding operator ($P_I$) to the renormalized Hilbert space,
$P_I=|\Psi_I^+\rangle\langle\Uparrow|+|\Psi_I^-\rangle\langle\Downarrow|$. Here, 
$|\Uparrow \rangle$, $|\Downarrow \rangle$ represent the renamed states 
$|\Psi_I^{+}\rangle$ and $|\Psi_I^{-}\rangle$, respectively
in the renormalized space to be considered as the new base kets.
$|\Psi_I^+\rangle$ and $|\Psi_I^-\rangle$ have the
following expression (for odd blocks) in the original spin Hilbert space
where $|\uparrow \rangle$ and $|\downarrow \rangle$ represent the eigenvectors of $S^z$ operator
at each site,
\begin{eqnarray}
|\Psi_I^+\rangle=a|\uparrow\uparrow\downarrow\rangle+ib|\uparrow\downarrow\uparrow\rangle
+c|\downarrow\uparrow\uparrow\rangle, \nonumber \\
|\Psi_I^-\rangle=a|\downarrow\downarrow\uparrow\rangle-ib|\downarrow\uparrow\downarrow\rangle
+c|\uparrow\downarrow\downarrow\rangle,
\end{eqnarray}
where
\begin{equation}
a=\frac{bD}{(\lambda-2 \varepsilon_0)},
b=\frac{1}{\sqrt{1+\frac{2D^2(4\varepsilon_0^2+\lambda^2)}{(4\varepsilon_0^2-\lambda^2)^2}}},
c=\frac{bD}{(\lambda+2\varepsilon_0)},
\label{abc}
\end{equation}
and $\varepsilon_0$ is the ground state energy of the block. The presence of bond alternation imposes 
to consider two types of blocks as depicted in Fig.\ref{blocks}, namely even and odd which are
the mirror image of each other. For even blocks we find similar eigenstates by replacing
$a \rightarrow -c$, $b \rightarrow b$ and $c \rightarrow -a$.

The global embedding operator is the direct product of the embedding operator of each block,
$P=\otimes_I^{N/3} P_I$ which gives the renormalized Hamiltonian by 
$H^{ren}=P^{\dagger} H P$ \cite{qrg}.
The renormalized Hamiltonian is similar to the original one, Eq.\ref{eq1}, while the coupling 
constants are replaced with the renormalized one (denoted with $'$ ) as given in the following equations,
\begin{equation}
\label{rgflow}
J'=X J, \hspace{8mm}  \lambda'=\lambda, \hspace{8mm} D'=-\frac{4ab^2c}{X} D,
\end{equation}
where $X=(1-2a^2)(1-2c^2)$.
These relations, Eq.\ref{rgflow}, define the QRG-flow of our model which will be
used in next sections and their features will be discussed in Sec.\ref{summary}.

\begin{figure*}[t]
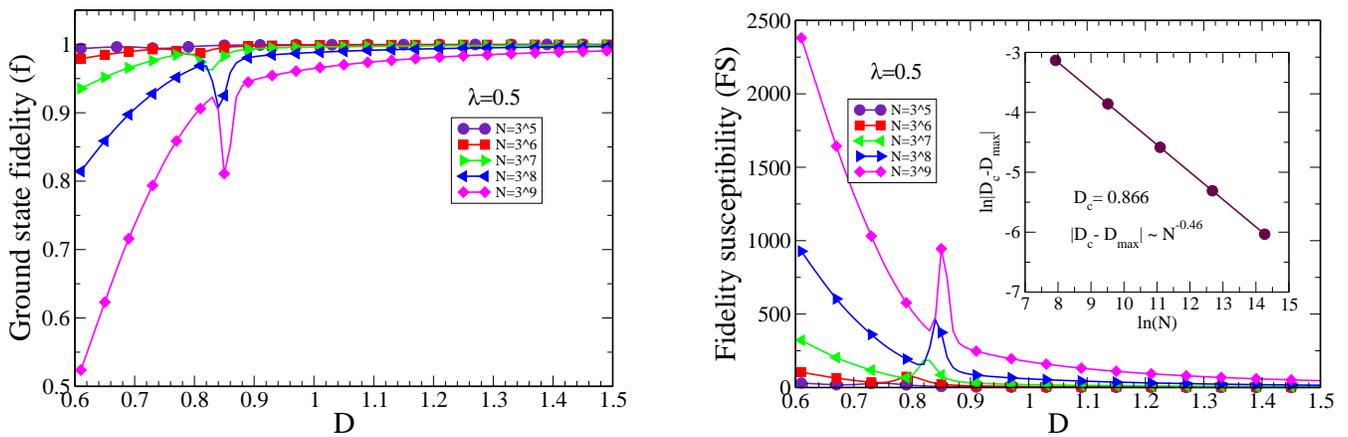
%
\includegraphics*[width=0.46\linewidth]{fidelity}
\hspace{10mm}
\includegraphics*[width=0.46\linewidth]{fs}
\caption{
(Left) Ground state fidelity (f) versus Dzyaloshinskii-Moriya interaction strength (D) for
different chain size and at $\lambda=0.5$.
(Right) Fidelity susceptibility (FS) versus $D$ at $\lambda=0.5$ for various system
sizes $N=3^5, \dots, 3^9$. Inset: the scaling behavior shows how the maximum of FS ($D_{max}$)
approaches the critical point ($D_c$) by increasing size. In both figures $\delta=0.01$ which 
is the difference between two parameters ($D_+-D_-=\delta$) to calculate fidelity in the left figure
and according to Eq.(\ref{eq12}) for the right figure.
}
\label{fs}
\end{figure*}

\subsection{Fidelity signature}
We implement the formalism introduced in \cite{Langari2012} to calculate the ground state
fidelity in terms of quantum renormalization group. 
According to the QRG-flow (Eq.\ref{rgflow}), $\lambda$ does not run within the QRG iterations which
hints to study the quantum phase transition by variation of $D$.
The fidelity (f) associated to the ground states $|\Psi(D)\rangle$ for a system of size $N$ is defined by
\begin{equation}
\label{fidelity}
f(D, \delta; N)=\langle\Psi(D_-)|\Psi(D_+)\rangle,
\end{equation}
where $D_{\pm}=D\pm \delta/2$ and $\delta$ is a small deviation around $D$.
According to the renormalization group approach $|\Psi\rangle=P|\Psi^{(1)}\rangle$ in which $P$ is 
the global embedding operator and $|\Psi^{(1)}\rangle$ is the ground state of the renormalized 
Hamiltonian. Thus, fidelity can
be written in terms of renormalized ground state,
$f=\langle\Psi^{(1)}(D_{-})|P^{\dag}(D_{-})P(D_{+})|\Psi^{(1)}(D_{+})\rangle$.
A straightforward calculation shows that
\begin{eqnarray}
\label{r0}
&P_{I}^{\dag}(D_{-})P_{I}(D_{+})=R_0(D_-, D_+) \mathcal{I}, \nonumber \\
&R_0=[a(D_{-})a(D_{+})+b(D_{-})b(D_{+})+c(D_{-})c(D_{+})], 
\end{eqnarray}
for both even and odd types of blocks
where $\mathcal{I}$ is the identity operator. Therefore, the first QRG iteration leads to 
$f=R_0^{\frac{N}{3}}(D_-, D_+)\times \langle\Psi^{(1)}(D_{-})|\Psi^{(1)}(D_{+})\rangle$,
where fidelity of the original model is expressed in terms of fidelity of the renormalized
one, i.e. $f=R_0^{\frac{N}{3}} f^{(1)}$. It defines the renormalization of fidelity in terms of QRG.
The QRG procedure is iterated $n$-times to reach the renormalized system of $N=3^{n+1}$ and
the ground state fidelity is expressed by
\begin{equation}
\label{frg}
f=(\prod_{i=0}^{n-1}R_i^{\frac{N}{3^{i+1}}})\langle\Psi^{(n)}(D_{-})|\Psi^{(n)}(D_{+})\rangle,
\end{equation}
where $R_i$ has the same expression as given in Eq.\ref{r0} for $R_0$ in which $a$, $b$ and $c$ are 
calculated at the $i$-th 
QRG iteration and $\langle\Psi^{(n)}(D_{-})|\Psi^{(n)}(D_{+})\rangle$ is the fidelity of a single block 
with three sites and $n$-times renormalized couplings.

We have plotted fidelity (Eq.\ref{frg}) versus $D$ in Fig.\ref{fs}-(left) for different chain length ($N$),
$\delta=0.01$ and
at $\lambda=0.5$. By definition, fidelity is bounded by $0\leq f \leq 1$ and an abrupt drop is a signature
of quantum phase transition. We observe a sharp drop in Fig.\ref{fs}-(left) for $0.8<D<0.9$ which manifests
that the ground state has encountered an essential change. The deep of drop is enhanced as the system
size is increased which justifies an unfailing drop in the thermodynamic limit. This signature of quantum 
phase transition is more pronounced in the fidelity susceptibility (FS) which is the leading nonzero
term in the expansion of fidelity and shows the change rate of fidelity, i.e. 
$f=1-\frac{\delta^2}{2} FS + O(\delta^3)$. Thus, FS is obtained by the following relation
\begin{equation}
\label{eq12}
FS=\lim_{\delta \rightarrow 0}2\frac{1-f}{\delta ^2}.
\end{equation}
Fig.\ref{fs}-(right) presents $FS$ versus $D$ for various system sizes, $\delta=0.01$ and at $\lambda=0.5$. 
A maximum appears in $D=D_{max}(N)$ which is increased by the size of system representing a
divergence in the thermodynamic limit. The position of $D_{max}(N)$ is exactly at the point 
where fidelity receives a drop.

\subsection{Scaling analysis}
It has been shown \cite{Gu2010} that the fidelity susceptibility at the quantum critical point
obeys a scaling relation. The scaling analysis for finite system size ($N$) states 
\begin{equation}
\label{scaling}
|D_c-D_{max}| \sim N^{-\frac{1}{\nu}},
\end{equation}
where $D_c$ is the quantum critical point, $D_{max}$ is the position of maximum in FS 
and $\nu$ is the critical exponent governing the divergence of correlation length. The analysis
of data of Fig.\ref{fs}-(right) is presented as an inset to this figure. It clearly verifies that the 
scaling relation Eq.\ref{scaling} is satisfied with $D_c\simeq0.866$ and $\nu\simeq2.17$ for $\lambda=0.5$.
Moreover, we have obtained similar behavior for different values of $\lambda$ (not presented here)
where the critical point is found to be a function of bond alternating parameter, 
$D_c \simeq \sqrt{1-\lambda^2}$ while $\nu\simeq 2.17$ is the same for all values of $\lambda$. 
In contrast to what stated in Ref.\cite{Hao2010} the critical exponent which we got does not
depend on $\lambda$. It means that the whole phase boundary for $0\leq \lambda <1$ belongs 
to the same universality class of a second order phase transition.

\begin{figure*}[t]
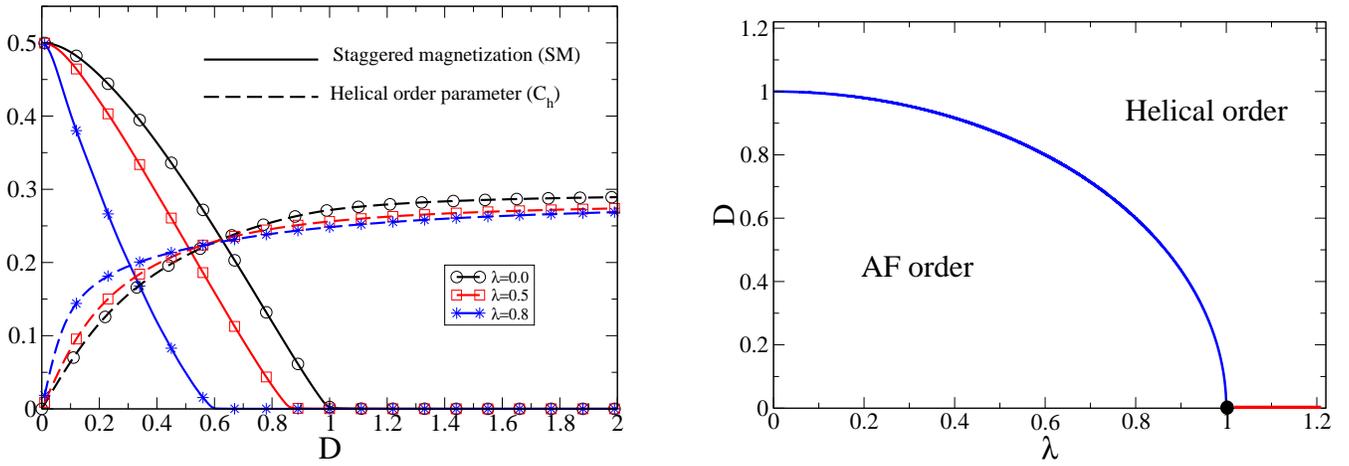
%
\includegraphics*[width=0.46\linewidth]{stmag-ch}
\hspace{10mm}
\includegraphics*[width=0.46\linewidth]{phase-diagram}
\caption{
(Left) 
Staggered magnetization (SM, solid lines) and helical order parameter ($C_h$, dashed lines) versus $D$ for 
$\lambda=0.0, 0.5, 0.8$.
(Right) Zero temperature phase diagram in $\lambda-D$ plane. The blue curve represent
the phase boundary between antiferromagnetic and helical ordered phases. The filled black circle denotes
the first order tri-critical point and the red line shows a ferro-antiferromagnetic phase.
}
\label{phase-diagram}
\end{figure*}

\section{Summary and discussions\label{summary}}
We have implemented the quantum renormalization group approach to study the zero temperature phase
diagram of bond alternating Ising chain in the presence of DM interaction.
To get a self similar Hamiltonian two types of blocks with 3-sites have been considered which leads
to the QRG-flow equations of Eq.\ref{rgflow}. The QRG-flow tells that $\lambda$ does not vary within
QRG procedure while $D$ is renormalized. To get the phase diagram of the model we have calculated
the renormalization of ground state fidelity which has been developed recently \cite{Langari2012}.

Fidelity as a geometric quantity \cite{zanardi2007} shows how much the ground state
encounters an essential change by slightly moving in the parameter space. Therefore, a sharp drop
of fidelity versus a control parameter is a signature of quantum phase transition. The renormalization
of fidelity obtained in Eq.\ref{frg} in addition to QRG-flow, Eq.\ref{rgflow} give the 
fidelity of our model for very large system sizes without the need to get the ground state.
A clear drop of fidelity versus $D$ in Fig.\ref{fs}-(left) and consequently a maximum in the
fidelity susceptibility, Fig.\ref{fs}-(right), verifies the existence of a quantum phase transition at $D_c$.
We have applied the finite size scaling on the susceptibility data presented in 
the inset of Fig.\ref{fs}-(right)
for $\lambda=0.5$ and generally got the critical phase boundary $D_c\simeq \sqrt{1-\lambda^2}$. 
The phase boundary which separates the antiferromagnetic (AF) N\'{e}el order
from a helical order is shown by the blue line in Fig.\ref{phase-diagram}-(right).
However, our scaling analysis gives a single value for the correlation length 
exponent $\nu \simeq 2.17$ (independent of $\lambda$) for the whole phase boundary which is 
in contrast to \cite{Hao2010}. Although the presence of bond alternation breaks the translation
invariance of the Hamiltonian it does not change the symmetry of the ground state which has 
already been spontaneously broken due to antiferromagnetic long range order.
A comparison of our results with \cite{Jafari2008} concludes that
the bond alternation does not change the universality class of the model as far as $\lambda \neq 1$.

We calculate the staggered magnetization (SM) and helical order parameter which is
presented in Fig.\ref{phase-diagram}-(left). Staggered magnetization is defined by 
\begin{equation}
SM=\frac{1}{N}\sum_{i=1}^{N} (-1)^i \langle \Psi| S_i^z |\Psi\rangle,
\label{sm}
\end{equation}
which can be expressed in terms of the renormalized ground state by replacing 
$|\Psi\rangle=P|\Psi^{(1)}\rangle$. We use the projection of spin operators into the
renormalized Hilbert space which finally leads to 
\begin{equation}
SM= - \frac{1-4b^2}{3} SM^{(1)},
\label{smrg}
\end{equation}
where $SM^{(1)}$ is the staggered magnetization of the renormalized chain. 
A large number of iterations of Eq.\ref{smrg}
give the staggered magnetization in the thermodynamic limit.
Similarly, the helical order parameter ($C_h$) is defined by the following relation \cite{Jafari2008}
\begin{equation}
C_h=\frac{1}{N} \sum_{i=1}^N \langle \Psi| (S_i^x S_{i+1}^y-S_i^y S_{i+1}^x) |\Psi \rangle,
\end{equation}
which can be calculated using the QRG-flow. In this respect, $C_h$ is expressed in terms of the helical 
order parameter in the renormalized model ($C_h^{(1)}$) by the following relation
\begin{eqnarray}
C_h=C_h^{(0)}&+& \frac{\Gamma^{(0)}}{3}C_h^{(1)}, \nonumber \\
C_h^{(0)}=\frac{b(a-c)}{3} \;\; &,& \;\; \Gamma^{(0)}=-4ab^2c.
\end{eqnarray}
The above equation is iterated by replacing the couplings with the renormalized ones to reach 
the stable thermodynamic limit.
We have plotted both staggered magnetization and
helical order parameter versus $D$ in Fig.\ref{phase-diagram}-(left) 
for three values of $\lambda=0.0, 0.5, 0.8$ and $N\rightarrow \infty$.
For $D=0$, $SM$ is at its saturated value $0.5$ and $C_h=0$. The onset of nonzero $D$ induces
a helical order on the spins in the presence of antiferromagnetic order. The increase of $D$ reduces the
antiferromagnetic order and enhances $C_h$ (helical order). Exactly at the quantum critical point $D_c$,
$SM$ vanishes and remains zero for $D\geq D_c$ while $C_h$ saturates to a finite amount which is 
less than its maximum attainable value. Similar behavior has been observed for all $0\leq \lambda <1$ while the 
saturated value of $C_h$ is slightly decreased by increase of $\lambda$.

To complete the phase diagram let us concentrate on the $D=0$ axis (Fig.\ref{phase-diagram}-(right)). 
At $\lambda=0$ the model
is simply an antiferromagnetic Ising chain with N\'{e}el order 
$|\uparrow\downarrow,\uparrow\downarrow, \cdots, \uparrow\downarrow\rangle$ and the ground state
energy is $E_0=-NJ/4$. For $0< \lambda <1$, there are two types of couplings $(1-\lambda)J$ and
$(1+\lambda)J$ which are positive and induce the previous N\'{e}el order and ground state energy.
Exactly at $\lambda=1$, the weak interaction $(1-\lambda)J$ becomes zero and the spin model
decomposes to $N/2$ independent pairs of antiferromagnetically coupled spins. The ground state energy is
still $E_0=-NJ/4$ while the ground state is exponentially degenerate, namely $2^{N/2}$. The degeneracy
arises from the pairs which are decoupled. For instance, 
$|\uparrow\downarrow,\uparrow\downarrow, \cdots, \uparrow\downarrow\rangle$,
$|\downarrow\uparrow,\uparrow\downarrow, \cdots, \uparrow\downarrow\rangle$
and $|\downarrow\uparrow,\downarrow\uparrow, \cdots, \downarrow\uparrow\rangle$ are examples of different
configurations which is possible for the ground state. 
The entropy (S) is proportional to $\ln (\# \mbox{of available states})$
which leads to $S\sim \ln(2^{N/2})=\frac{N}{2}\ln 2$.
This high amount of entropy at $\lambda=1$ 
is a signature of a phase transition which is actually of first order. Meanwhile, for $\lambda>1$ one of
the interactions becomes ferromagnetic, $(1-\lambda)J<0$ and the other $(1+\lambda)J$ is still 
antiferromagnetic which totally leads to $E_0=-N \lambda J /4$. Thus, the derivative of $E_0$ with 
respect to $\lambda$ receives a discontinuity at $\lambda=1$. The ground state for $\lambda>1$ 
is either $|\uparrow\downarrow,\downarrow\uparrow,\uparrow\downarrow,\downarrow\uparrow, \cdots \rangle$
or $|\downarrow\uparrow,\uparrow\downarrow,\downarrow\uparrow,\uparrow\downarrow,  \cdots \rangle$.
The first order tri-critical point is represented by the filled black circle in 
Fig.\ref{phase-diagram}-(right) and
the ferro-antiferromagnetic ground state is denoted by the red line for $\lambda>1$.

{\bf Acknowledgments} We would like to thank A. T. Rezakhani for fruitful discussions.
This work was supported in part by
Sharif University of Technology's 
Center of Excellence in Complex
Systems and Condensed Matter and the Office of Vice-President for Research. 
A. L. acknowledges partial support from the Alexander von Humboldt Foundation
and Max-Planck-Institut f\"ur Physik komplexer Systeme (Dresden-Germany).

\end{document}